\documentclass{cjaa}    
\usepackage{amsmath}
\usepackage{graphicx}
\usepackage[a4paper]{hyperref}
\input{psfig.sty}                       %for PS/EPS graphics inclusion, old
\usepackage{afterpage}

\def\lesssim{\mathrel{\mathchoice {\vcenter{\offinterlineskip\halign{\hfil $\displaystyle##$\hfil\cr<\cr\sim\cr}}}{\vcenter{\offinterlineskip\halign{\hfil$\textstyle##$\hfil\cr <\cr\sim\cr}}} {\vcenter{\offinterlineskip\halign{\hfil$\scriptstyle##$\hfil\cr <\cr\sim\cr}}} {\vcenter{\offinterlineskip\halign{\hfil$\scriptscriptstyle##$\hfil\cr <\cr\sim\cr}}}}}
\def\grtsim{\mathrel{\mathchoice {\vcenter{\offinterlineskip\halign{\hfil $\displaystyle##$\hfil\cr>\cr\sim\cr}}}{\vcenter{\offinterlineskip\halign{\hfil$\textstyle##$\hfil\cr >\cr\sim\cr}}} {\vcenter{\offinterlineskip\halign{\hfil$\scriptstyle##$\hfil\cr >\cr\sim\cr}}} {\vcenter{\offinterlineskip\halign{\hfil$\scriptscriptstyle##$\hfil\cr >\cr\sim\cr}}}}}

\begin{document}
\def\teff{$T\rm_{eff }$}
\def\kms{$\mathrm {km s}^{-1}$}

\title{The\ Local-Galactic interpretation of the Gamma-Ray Bursts}
\author{Wolfgang Kundt}
\institute{Argelander Institute for Astronomy of Bonn University, Auf dem H\"{u}gel 71, D-53121 Bonn, Germany}
%\author{WOLFGANG\ KUNDT\\Argelander Institute of Bonn University, Auf dem H\"{u}gel 71, D-53121 Bonn, Germany}
\abstract{
In this talk, I shall update my 16-year old claim that all the (thousands of)
observed GRBs - both long and short, repeating or (so far) not - come from the
surfaces of Galactic neutron stars, often called 'magnetars', or 'throttled pulsars'.
\keywords{gamma-ray bursts, magnetars, afterglows, hosts}
}
\titlerunning{Gamma-Ray Bursts}

\maketitle

\section{Introduction}

A fresh attempt is made to convey the message that all the detected and
catalogued $\gamma$-ray bursts (GRBs) come from nearby Galactic neutron stars
--  rather than from cosmologically distant sources of similar type\ -- as has
already been thought during the 1980s. This conviction is sustained by their
otherwise gigantic energies radiated at almost microscopic time scales, and
their occasional extremely hard spectra, reaching and exceeding TeV energies.
Relativistic redshifts stem not only from cosmic distances, but can likewise
be generated by nearby Galactic neutron stars.

\section{The GRBs, their afterglows, and their distances}

Here at Vulcano, I have repeatedly talked about GRBs coming from nearby
Galactic neutron stars, at distances between 10 and 500 pc, preferentially
between 100 and 200 pc (Kundt, 2006, 2008a, 2009), so I will skip the
historical part this time. I still like the generally accepted
Galactic-neutron-star interpretation from the 80s, but not its later
`improvements' which were suggested by (i) the high isotropy of their
celestial distribution, (ii) their cosmologically large redshifts, (iii) their occasional
host galaxies, and (iv) their occasional (though rare) supernova-like appearances.

\afterpage{\clearpage}
\begin{figure}[t]
\begin{center}
%\resizebox{\hsize}{!}{\includegraphics[clip=true]{kundt_2009_01_fig01.eps}}
\resizebox{9.0cm}{!}{\includegraphics[clip=true]{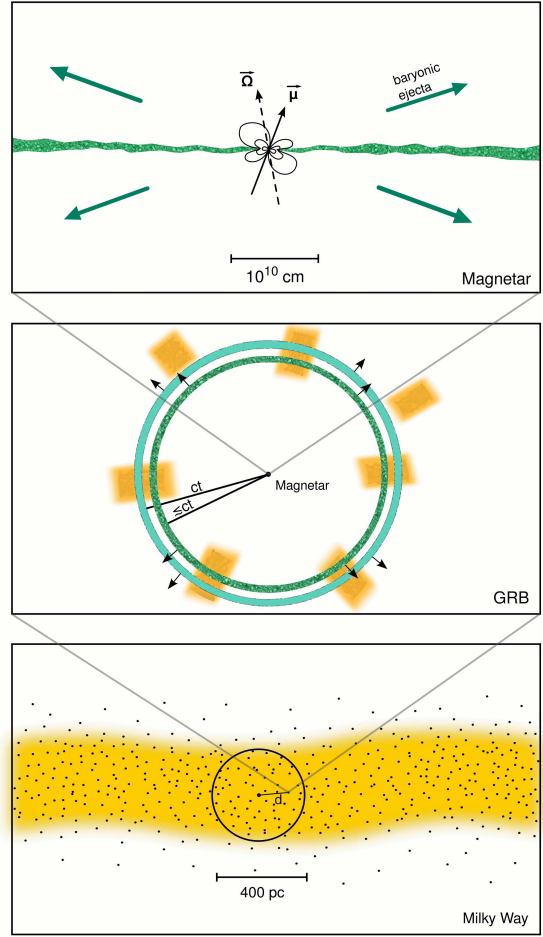}}
\end{center}
\vspace{-0.9cm}
\caption{
Sketch of the (preferred, nearby Galactic)\ source morphology of all
the GRBs. (a) The corotating magnetosphere of a throttled Galactic pulsar, or
magnetar -- usually of age some $10^{6.4}$\ years, spin period between $5$ and
$12$ sec\ -- is strongly indented by its low-mass accretion disk, of typical
mass some $10^{-5}M_{\odot}$. A GRB\ is emitted when a large chunk of
(decelerated) matter, of mass some $10^{15}$g , falls down upon its surface,
from the inner edge of its (throttling) disk, and part of the chunk's matter
is centrifugally re-ejected at transrelativistic speeds, across
speed-of-light-cylinder distances, of order $10^{10.5}$cm. In this scenario,
ions and electrons can be boosted to energies $\Delta W=e\int(\overset
{\rightarrow}{E}+\overset{\rightarrow}{\beta}\times\overset{\rightarrow}%
{B})\cdot$ $d\overset{\rightarrow}{x}$ $=$ $10^{21}$eV$(\beta_{\perp}%
B)_{12}(\Delta x)_{6.5}$ .\ (b) The disturbed magnetar emits a quasi
spherical, $\gamma$-ray-hot electromagnetic flash (blue) closely followed by a
trans-relativistic baryonic flash (green), both of which interact with the
magnetar's CSM (of individual morphology, drawn yellow), and radiate the
burst's afterglow. This scheme of interaction is thought to hold for all types
of GRBs, both the short and the long ones, and also the SGRs (which are
nearest to us). (c) These sporadic $\gamma$-ray bursters are comparable in
number to the (Galatic) pulsars, in total some $10^{7}$, and form an almost
%spherical distribution around the Sun for distances $\lesssim 0.3$\ kpc,
%spherical distribution around the Sun for distances \mbox{$\lesssim 0.3$}\ kpc,
spherical distribution around the Sun for distances \mbox{$\stackrel{<}{\sim} 0.3$}\ kpc,
whereby deviations from strict isotropy should increase with increasing
distance, hence with decreasing brightness.
}
\label{Fig.1}
\end{figure}

Item (i), the high `isotropy', clearly asks for either a very near, or else a
very far source population. Guided by their energetics, my preference is for
the former: The sources fit nicely into the Milky-Way disk, with typical
distances $\lesssim10^{2.3}$pc, see Fig.1c. Why do we not observe a strong
enhancement in directions of the Milky-Way disk? Not only because the farther
the sources the dimmer, but also because the old neutron stars' low-mass
accretion disks -- assembled from the ISM -- tend to be oriented
preferentially at right angles to the Galactic disk (through which all stars
oscillate). Their individual emissions are expected to be mildly beamed w.r.t.
their disk planes, hence their superposition slightly beamed perpendicular to 
the Milky-Way disk, as
already quantified in Kundt \& Chang (1993). Note that with increasing
sensitivity and duration of our surveys, the log(N)-log(S) diagram has
increasingly evolved towards a (Euclidean) $S^{-3/2}$ distribution whose lower
and upper turnover intensities $S_{\min}$ and $S_{\max}$ have moved apart from
each other with observation epoch, corresponding to a growing distance ratio
$d_{\max}$/$d_{\min}$ of the detected sources -- because of $\ S$ $\sim
d^{-2}$\ -- from an initial $2$ to a later $10$, see Fig.13 in Fishman \&
Meegan (1995). There are many more faint bursts in the sky than were initially
known, whose distribution may well map the (near part of the) Milky-Way disk. \ 

%\afterpage{\clearpage}
\begin{figure}[t]
\resizebox{\hsize}{!}{\includegraphics[clip=true]{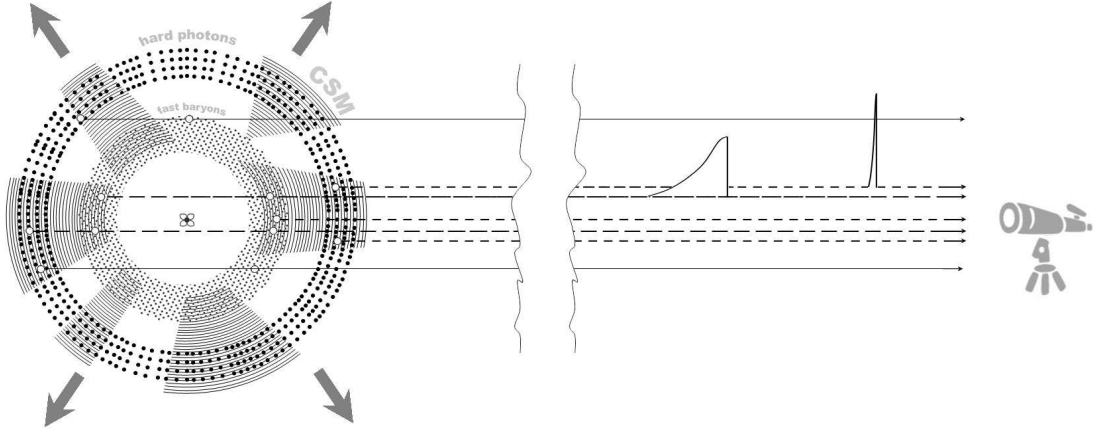}}
\caption{
Simplified Sketch of my understanding of a GRB's geometry: When a
(terrestrial-mountain-sized) chunk of matter hits the surface of a neutron
star, a \textit{hard photonic} GRB is emitted, followed (within seconds) by a
burst of transrelativistic baryons (from the strongly heated, accreted matter,
ejected centrifugally). This pair of (almost) luminally expanding shells
impacts on the ambient \textit{CSM}, causing it to radiate. A distant observer
sees first partially blueshifted radiation, (soon) thereafter partially
redshifted radiation from the shells' approaching and receding hemispheres.
The figure represents blue- and red-shifted photons by short- and long-dashed
straight lines, unshifted photons by unbroken lines. Blueshifted emissions and
absorptions have not yet been recorded, whilst the detected
\textit{red-shifted absorptions} (of the afterglow) should stem from the left
hemisphere of the \textit{fast baryonic} shell when it is crossed by emission
from its preceding hard photonic shell, on interaction with the CSM, (5th ray
from above). \textit{Red-shifted emission} is expected from the left
hemisphere when the baryonic shell crosses some CSM, (3rd ray from above).
(Strong, aligned) blue-shift and red-shift obey: $z+1$ = $\gamma$(1+$\beta$).
Whilst strongly blue-shifted emission will reach the observer in the form of
early, sharp spikes, strongly red-shifted emission will arrive more uniformly,
throughout epochs between seconds and years, largely dependent on the
distribution of the CSM around the central neutron star.
}
\label{Fig.2}
\end{figure}

Item (ii), the `cosmological' redshifts of their afterglows, are often
variable (Greiner, 2008), and do not correlate with distance (Song, 2008); in
particular, the implied time-dilation of high-z bursts is absent in the data
(Crawford, 2009). Instead, the observed redshifts can be understood as purely
kinematic, reaching us from the far side of an expanding shell of baryonic
ejecta, whilst their (bright) blueshifted analogues reach us only during
short-lasting onsets of subbursts (FREDs), so far undetectably fast, cf.
Figs.1b and 2. Observed redshifts of $z\lesssim9$ correspond to Lorentz
factors $\gamma\lesssim5$, according to the 1-d Doppler formula
\begin{equation}
z+1=\gamma(1+\beta)\mbox{ \ .}%
\end{equation}
We deal with transrelativistic ejecta of slowly spinning neutron stars --
dying (or `throttled') pulsars, often called \ `magnetars'\ (Kundt, 2008a) --
which impact onto their circumstellar medium (CSM) and cause it to flare,
reminiscent of supernova (SN) ejecta, though less massive, and faster.

Item (iii), the often reported distant `host galaxies', is plainly
inconclusive. Originally, there was the ''no-host dilemma'' of Brad Schaefer
(1999) which demonstrated an anti-correlation between GRBs and large galaxies.
None of the (massive) catalogued galaxies have ever served as a host for a
GRB; the published hosts form a heterogeneous set of faint, low-mass, ''very
peculiar'' luminous objects (Savaglio et al, 2007). Some 50\% of the proposed
(50\%) hosts of long GRBs ($\Delta t$ $\grtsim$ 2s) may even be ''chance
superpositions'' (Cobb \& Bailyn, 2007), so that only some 25\% of all bursts
have confirmed hosts. GRB 070125, a long burst, has been called a ''shot in
the dark'' by Robert Naye and Neil Gehrels, because no host could be found for
it in a clear sky. I like to think of the (well established)\ `hosts' as light
echos, or transient reflection nebulae, to be considered below. The
untenability of the host interpretation gets evident, in a large number of
cases, via a simple energy balance: If a burst injected an (electromagnetic)
energy of order $10^{54\pm1}$erg, or even $10^{55\pm1} $erg (near TeV
energies: Atkins et al, 2003), into a host galaxy of luminosity $\lesssim
10^{43}$erg/s (Savaglio et al, 2007), a significant percentage $\eta$\ of its
injected power would be subsequently re-radiated by the burst's CSM, and be
visible as a flaring (unresolved) afterglow point source of power
$\lesssim10^{48}\eta_{-2}t_{6}^{-1}$erg/s during the first few weeks and
months $t$\ after the burst, with $t_{6}/\eta_{-2}$ $\grtsim1 $. This
afterglow brightness would be strongly variable, controlled by the structure
of the burst's CSM, and exceed the host's brightness (bolometrically) by some
five times $\eta_{-2}t_{6}^{-1}\approx5$\ orders of magnitude! Nothing like
this has ever been seen.

Item (iv), a coincidence with distant `supernovae', has been reported for
four or more bursts of long duration ($\Delta t$ $\grtsim$ 2s) with lowest
redshift z ($\lesssim$ 0.2), cf. Bloom et al (1999). Their afterglows show a
bump in the optical lightcurve looking like that of a SN, and even their
optical spectra look SN-like, of (special) type Ib or Ic. Instead of realizing
a power-law decrease, their afterglows wane exponentially (for at least one
month). Do we deal with coincidences between a SN and a GRB? \ Such a
coincidence of explosions --\ of which we witness less than 5 per day in the
whole Universe in the case of the GRBs, and less than 2 per week in the case of
SNe -- do not have a realistic a-priori chance of coinciding once per century
in the sky unless they were causally connected. But a GRB lasts typically
$\lesssim$ one minute whilst the first light from a SN comes from a sphere of
radius several light-minutes --\ when its piston reaches the progenitor star's
outer edge, and launches a UV flash -- so that coincidences with GRBs are not
at all expected; (a first recorded case was SN 2008D, Soderberg et al 2008).
In my understanding, a SN creates neither a GRB, nor a jet (Granot 2007;
Kundt 2008b). A coincidence with a GRB could therefore only happen by chance
projection, whose probability is zero. On the other hand, in my understanding
of GRBs, the impacted neutron star (by a clump of matter from its inner
accretion disk) ejects a baryonic shell at transrelativistic speeds,
centrifugally -- the slower and the thicker the shell the longer are the burst
and its early afterglow -- and the smaller is its redshift z. Precisely this
correlation has been found: Only long-duration bursts of small z have had SN-like
afterglows. The physics of a GRB is not all that different from that of a
(core-collapse) supernova: In both cases, a central explosion ejects a
filamentary shell of processed baryonic material at high velocities. The GRBs
form the low-mass, high-velocity tail in this distribution, and the slowest
among them have SN-like afterglows. Again, Galactic neutron stars qualify as
their sources; no SN is required. Note that two counter examples to SN-like
afterglows of low-z, long GRBs have been discussed by McBreen et al (2008);
see also Bisnovatyi-Kogan (2006).

Why can friends of mine be convinced, nevertheless, of the occasional supernova-GRB
association? Let me expound on a recent best-case example, published by
Sonbas et al (2009). By routinely interpreting the afterglow redshift z (= 0.0331)
of GRB 060218 as measuring cosmic-recession speed -- rather than the outburst 
velocity from a Galactic explosion -- the authors feel urged to interpret its 
strong Balmer emission lines as due to an energetic (and focussed!) Hubble flow 
inside a "relic wind envelope around a core-collapse progenitor star" of a 
distant SN, of forbiddingly large energy, and of a forbiddingly long 
"shock-breakout time" of $\grtsim$\ 10 h (rather than $\lesssim$\ 1 h; cf. 
Colgate (1968), or rather Kundt (2005,2008b)). Spectra can allow for 
alternative interpretations.

The main reason that makes me mistrust the proposed cosmologically-far
interpretations of the GRBs is their exotically large implied powers, from
sources like neutron stars and/or BH candidates of which we have hundreds of
well-studied representatives in our Milky Way. All the Galactic copies respect
(approximately) the Eddington limit for a neutron star, $L_{Edd}=10^{38.3}$erg
s$^{-1}$($M$$/1.4M_{\odot}$) \ -- \ except for short-time outbursts which
violate the isotropic-feeding assumption in the derivation of the Eddington
constraint, and exceed $L_{Edd}$ by factors of $\lesssim10^{3}$ -- \ whilst
their distant brothers would have to transiently shine at ($d_{2}/d_{1}%
$)$^{2}\approx10^{16}$ times that much (for respective distances of
$d_{j}$ $\approx$\{$10^{10},10^{2}$\}pc). They would form a disjointly
different class of sources, of which there is no single local representative.
For a few years, the excess-energy factor $10^{16}$ has been lowered by model
builders -- by a factor of $10^{-4}$ to $10^{-6}$\ -- via a beaming hypothesis
(for the prompt emission) which can no longer be maintained once a continuity
has been established between `prompt' and `afterglow' emissions, cf.
Chincarini et al (2006), Kann (2008) for GRB 080319B, Romano et al (2006) for GRB 060124, GRBs 990123, 060729, and Fig.3.

\begin{figure}[t]
\resizebox{\hsize}{!}{\includegraphics[clip=true]{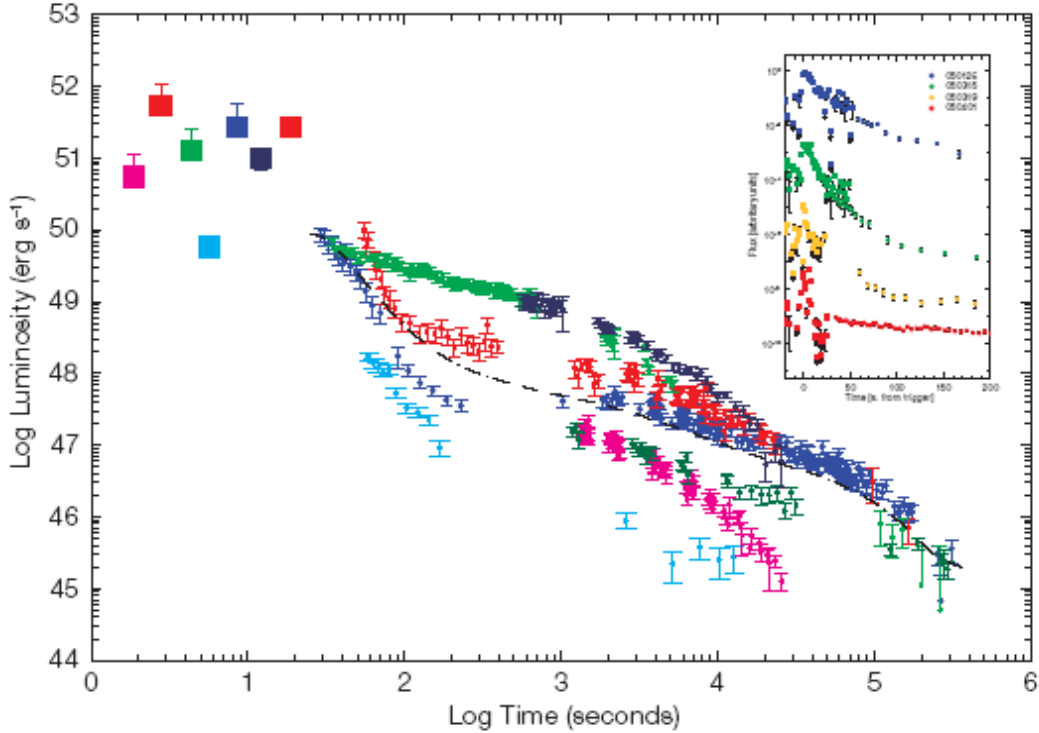}}
\caption{
Quasi-bolometric lightcurves (in colour) of seven\ GRBs and their
afterglows, log $L$ vs log $t$, with initial temporal gaps of \mbox{$\stackrel{>}{\sim} 30s$},
copied from Chincarini et al (2006). The data suggest -- as has meanwhile been
multiply verified -- that the (hard) prompt emission changes continuously into
the (softer) afterglow emission, or rather: that intensity-wise, there is no
well-defined decomposition into two disjoint emission modes.
}
\label{Fig.3}
\end{figure}

Before presenting the details of my `throttled-pulsars' model, and indicating
its viability, let me recall a number of distance estimates which support the
local-Galactic interpretation. The first stringent distance estimate was
published by Schmidt (1978), a year before the giant burst GRB 790305 (which
projects onto N49 in the LMC, as the first soft $\gamma$-ray repeater (SGR)).
Schmidt's estimate was soon followed by Aharonian \& Ozernoy (1979), also by
Zdziarsky ($\leq$1984), and by Colgate \& Petschek (1981). The estimate uses
the condition that near the source (of restricted surface area), hard photons
(above MeV) would pair-produce, and thus downscatter the spectrum towards
softer energies. For the maximal source distance $d$, it can be written as:
\begin{equation}
d<\mbox{kpc }/\mbox{ }\sqrt{S_{-4}}\label{prompt}%
\end{equation}
where $S$ is the burst's maximal energy flux at $\grtsim$\ MeV energies (and
$S_{-4}:=S/10^{-4}$cgs units, as always). This estimate excludes a
location of the burster in the LMC. Baring (1992) has stressed the assumption
in this inequality that the hard photons should not be emitted in a perfectly
aligned manner, where ''perfect'' would mean $\lesssim10^{-3}$ for GRB 790305,
and $\lesssim10^{-8}$\ for the non-SGR bursts. For a thermal emitter, such an
assumption sounds overly conservative to me; it was repeatedly discussed by
Zdziarsky. With this estimate, I strongly disagree with more recent authors
like Lithwick \& Sari (2001) who are happy to violate above inequality
(\ref{prompt}), and postulate a new type of jet formation, different from that
of the well-observed ones (Kundt \& Krishna 2004). Note that the estimate
(\ref{prompt}) differs from redshift-based estimates by a typical factor of
$10^{8}$.

A distance estimate similar to (\ref{prompt}) was presented during the 80s,
based on neutron-star energetics:
\begin{equation}
d<\mbox{kpc }\gamma\mbox{ }\sqrt{L_{38}}\label{Eddington}%
\end{equation}
in which $\gamma$\ is the (possible) bulk Lorentz factor of the
emission, and $L$ its power emitted by the source. Larger distances require
more powerful engines.

Now comes an estimate which I have not yet seen in publications by others,
even though it is at least as basic as the two preceding ones:
\begin{equation}
d<0.3\mbox{ kpc }/\mbox{ }\sqrt{S_{-10}}\label{afterglow}%
\end{equation}
in which this time, $S$ $:=$ $\nu S_{\nu}$ denotes the early
afterglow's energy flux density at frequency $\nu$ right\ after onset, which
tends to be almost frequency-independent (at about a fortnight after the
outburst), between mm wavelengths and X-ray energies. For this estimate, the
assumption is made that the (early!) afterglow is emitted incoherently, hence
governed by Planck's law. At visible frequencies $\nu$, the estimate is
obtained by comparison with the flux $S_{\odot}\approx10^{6}$erg
cm$^{-2}$s$^{-1}$ which we receive from the Sun (at 1 AU distance), whose
luminous area $\pi R_{\odot}^{2}$ can at best be available to the emitted
flash for times $t>$ 2s 
%TCIMACRO{\TEXTsymbol{>}}%
%BeginExpansion
%>$%
%EndExpansion
after onset, when the flash has reached a (radial) separation $R=ct$ $\geq
R_{\odot}$\ from the burst center. For lower than visible frequencies, the
Rayleigh-Jeans law $\nu L_{\nu}=4\pi R^{2}\sigma T^{4}(\nu/\nu_{peak})^{3}$
(for emission at frequencies below the peak frequency $\nu_{peak}$), when
combined with the flux propagation law $L=4\pi d^{2}S$ at distance $d$, yield
a similar distance estimate at slightly later times -- depending on the
detailed circum-burst medium\ -- whereby the onset time $t$ of the afterglow
at wavelength $\lambda$ should obey
\begin{equation}
t\geq10^{4.8}\mbox{s }\lambda_{-1}^{3/2}\mbox{ \ .}\label{time}%
\end{equation}
\ This onset time $t$\ ranges in minutes for IR wavelenths, and in days for
mm wavelenths, in reasonable agreement with the (few) observations (e.g. of
GRB 090423, Greiner 2009). The earlier the afterglow is caught, and the lower
the frequency $\nu$\ at which it is caught (at fixed $S$), the tighter is the
distance constraint. Note that the brightest ever optical burst, GRB 080319B,
could have been seen with the naked eye, at 5.3 mag, simultaneously with its
prompt $\gamma$-ray emission! For such a feat, the surface area of a neutron
star at a cosmical distance is largely insufficient, even if glowing at hard
X-ray temperatures. Future observations can still sharpen this estimate, down
to distances of \ $\grtsim10$ pc. But by this time it should have become clear
that cosmic distances of the bursters would violate fundamental physical constraints.

The ($\grtsim$8) SGRs tend to be judged at Galactic distances -- with the
possible exception of the 5 March 79 event (which projects onto N 49 in LMC,
cf. distance estimate
(\ref{prompt})) -- even though their rare giant bursts are indistinguishable
from ordinary GRBs. They are: SGR 0526-66, 1806-20, 1900+14, 1627-41 (Woods et
al 1999), XTE J1810-197, SWIFT J1955+2614, GRB 070610 (Castro-Tirado et al
2008, Stefanescu et al 2008), and GRB 070201. Note that the last-two listed
SGRs have also been viewed as GRBs of redshift $z=$ $0$. In my understanding,
they are the nearest among all GRBs, from whom we even see the many softer,
fainter repetitions, down in energy by a factor of order $10^{-3}$; cf.
Wachter et al (2007, 2008). For them follow three more distance estimates. The
angular speed of SGR 1806-20, and expansion speed of its newly created radio
bubble, during its giant outburst on 27 December 2004, were so large that
\begin{equation}
d\lesssim30\mbox{ pc }\beta_{-2.9}\label{speed}%
\end{equation}
must hold for familiar Galactic proper-motion speeds $\beta:=v/c\lesssim
10^{-3}$. This same estimate obtains when its maximal power (at outburst) is
postulated to conform with the (five-thousandfold weakened) Eddington limit
$L\lesssim10^{42}$erg/s (for clumpy feeding):
\begin{equation}
d\lesssim30\mbox{ pc }\sqrt{L_{41.9}}\mbox{ \ .}\label{power}%
\end{equation}
The ($\lesssim 10^{2.8}$ times) larger distance estimates favoured in the
literature were all suggestive but not conclusive (Kundt 2006). Finally, the
Cavallo-Fabian-Rees limit (on $\Delta L/\Delta t$) conforms with the latter
estimate (Vietri et al 2007), as an independent fundamental constraint.

For completeness' sake, here are another twenty indications against the
cosmological-distance interpretation of the GRBs, taken mostly from (Kundt,
2009):

$\bullet$ (j) The X-ray afterglow of GRB 031203 was resolved into $\grtsim$~2
expanding (noisy) rings, of radii $\lesssim$$3^{\prime}$, during 10 successive
10$^{2}$min-observations, starting 6 hours after the burst, explained via
(Galactic) foreground scattering (Vaughan et al 2004, also: Kundt 2009).

$\bullet$ (jj) There are frequent precursor events to GRBs, up to $10$ minutes
at least (Burlon et al 2008, Romano et al 2006), and postcursor events, of
temporal offset $\lesssim1$ hour, both of (smaller but) comparable total
energy; (cf. Wang \& M\'{e}sz\'{a}ros 2007, who restrict their discussions to
offsets by $\lesssim10^{2}$s). Such temporal clusterings of supposedly quite
rare, gigantic explosions pose problems to the cosmological interpretation.

$\bullet$ (jjj) GRB 060729 had an X-ray afterglow that hardly faded for 125
days\ (Grupe et al 2007). A similar case may have been GRB 070110 (Troja et al 2007).

$\bullet$ (jv) GRB 030329 showed (supposedly) two superluminal expansions, at
($4\pm1$)c, and at $19$ c, (Taylor et al 2004, 2005). Cf. the giant outburst
of SGR 1806-20 on 27 Dec. 2003, whose radio bubble (supposedly) expanded transluminally.

$\bullet$ (v) Afterglow brightnesses are $z$-independent (Vreeswijk et al 2004).

$\bullet$ (vj) The X-ray-afterglow spectra do not reveal the expected increasing
degree of ionization of their CSM into which their bursts should penetrate.

$\bullet$ (vjj) The lightcurves of the X-ray afterglows show strong flares
($\lesssim10^{2}$), between minutes and days after outburst, as well as steep
breakoffs (Chincarini et al 2007).

$\bullet$ (vjjj) GRBs show brightness excesses at the high-$z$ end (Schaefer 2007).
Their inferred luminosities, hardnesses, and variabilities grow like powers of z 
(Yonetoku et al, 2004, Graham et al, 2009).

$\bullet$ (jx) GRBs show hardness excesses at $\lesssim$ $10$ TeV (Atkins et al 2003).

$\bullet$ (x) GRBs show occasional duration excesses, of (even) $\grtsim$ hour
(Fishman \$ Meegan 1995, Fig.8).\ 

$\bullet$ (xj) None of the bursts has ever shown a long-dis$\tan$ce travel
signature (Mitrofanov 1996).

$\bullet$ (xjj) No orphan afterglows have ever been detected, (i.e. afterglows
whose generating bursts were beamed away from us) (Rau et al 2006).

$\bullet$ (xjjj) GRB 070201 has not been seen at gravity waves (by LIGO),
against expectation (Svitil 2008).

$\bullet$ (xjv) The accreting Galactic dead-pulsar population should be
detected, at a generally agreed integrated mass rate of $10^{-17}$M$_{\odot}
$/yr n* (Kundt \& Chang 1993).

$\bullet$ (xv) The so-called host galaxies have (atypically low) luminosities:
\ $L/L_{\odot}$ $\epsilon$ ($10^{7}$,$10^{10}$), (Savaglio et al 2007). See
also Schaefer (2006) for host problems with the short GRBs.

$\bullet$ (xvj) Three-colour plots of optical afterglows do not show a large
scatter, and signal a two-temperature (!) structure of their sources.

$\bullet$ (xvjj) Mi$\lg$rom \& Usov (1995) have presented (weak) evidence for a
common origin of UHE CRs and GRBs, via their occasional (almost) coincident
positions in the sky, and (weakly) correlated TOAs. (Note that UHE ions
propagate almost like photons, except for slightly curved orbits in the
Galaxy's magnetic fields, of curvature radius $R_{B}=\gamma m_{0}c^{2}
\beta_{\perp}/ZeB$ $=$ 2 kpc $\gamma_{10}(m_{0}/m_{p})/ZB_{-5.3}$ ).
This suggestion agrees with mine, except that in my understanding, both source
classes are local Galactic (Kundt 2005, 2009).

$\bullet$ (xvjjj) The column densities of the damped Lyman $\alpha$ systems
(DLAs) in GRB spectra\ are either larger, or else smaller than in most quasar
spectra (Vreeswijk et al 2004).

$\bullet$ (xjx) The (strong) Mg II absorbers in their spectra are 4-times
overabundant w.r.t. quasars, and variable on the timescale of hours (Sudlovsky
et al 2007).

$\bullet$ (xx) The log($N$) vs log($S$)-distribution of the GRBs signals a
thick-shell distribution in power, with
%TCIMACRO{\TEXTsymbol{<}}%
%BeginExpansion
$<$%
%EndExpansion
$d_{\max},d_{\min}>$
%TCIMACRO{\TEXTsymbol{>}}%
%BeginExpansion
%$>$
%EndExpansion
$\grtsim$ 5 (Fishman \& Meegan 1995 Fig.13, Pendleton et al 1997); whereas
log($N$) vs log($L$) ranges through many orders of magnitude, cf. (vjjj).

\section{The Sources of the GRBs}

In this section I shall update my 1993 model (with Hsiang-Kuang Chang), in
which the bursters are assumed to be (mostly) the dead-pulsar population:
nearby Galactic neutron stars, whose wind-blown cavities have collapsed after
some $10^{6.4}$years (of pulsar life), when their spin period had grown
towards the $5$ to $12$ seconds interval (of the dying pulsars), and a
low-mass accretion disk has formed around them which indents into their
corotating magnetospheres, as sketched in Figs.3a,b of (Kundt 2009). Many of
these\ `throttled pulsars' are observed as soft X-ray sources, among them the
anomalous X-ray pulsars (AXPs), soft gamma-ray repeaters (SGRs), recurrent
radio transients (RRATs), or `stammerers', or `burpers', and the `dim isolated
neutron stars' (DINSs), with the following properties (Kundt 2008a):

The dying pulsars are isolated neutron stars, with spin periods $P$\ between
$5$s and $12$s, and similar glitch behaviour to other neutron-star sources.
They are soft X-ray sources, hotter than pulsars of the same spindown age by a
factor of $\grtsim$ $3$, mostly without pulsed coherent radio emission.
Their spindown is rapid, $\tau$ = $10^{4\pm1}$yr, despite ongoing accretion.
Their estimated number in the Galaxy is large, comparable to the number of
pulsars, but due to their short spindown times, of order $10^{4}$yr \ --
compared with their age, of order $10^{6.4}$yr \ -- their detectable number in
the sky is reduced by a factor of $10^{-2.4\ \pm\ 0.5} $ compared
with ordinary pulsars. Most of their power is derived from accretion, whose
implied (small) spinup is overcompensated by magnetospheric spindown. They are
often (some 50\%) found near the center of a pulsar nebula.

Throttled pulsars cannot only form from dying pulsars, as just explained, but
also from newborn pulsars, via `fallback matter', right after their birth
inside a SN shell, when a small fraction of the ejected matter does not make
its way to infinity. Such young pulsars can be seen embedded in an X-ray
nebula, whose innermost portion may well indent into the pulsar's corotating
magnetosphere. Again we deal with a (mildly) throttled, magnetized, spinning
neutron star.

This rather inconspicuous class of throttled pulsars can do two further
things: It can emit cosmic rays, quasi-steadily, preferentially from the
disk's inner edge to which the corotating magnetosphere is stick-slip coupled.
And it can flare in the form of an impressive burst of gamma-rays through
X-rays whenever a large chunk of (disk) matter gets sufficiently decelerated,
via recoil on the (tangentially) ejected cosmic rays, and falls down onto the
neutron star's surface, liberating its huge gravitational potential, and
heating up to temperatures $T$ of order
\begin{equation}
T\lesssim\ G\mbox{ }M\mbox{ }m\mbox{ }/\mbox{ }R\mbox{ }k=10^{12.2}\mbox{K
}(m/m_{p})\mbox{ \ .}\label{temperature}%
\end{equation}
Such dumped matter will cool immediately, via neutrino and photon radiation,
via energy-sharing with crustal matter, and via rebounce and adiabatic
expansion, to heights exceeding the neutron star's radius $R$ = $10^{6}$cm. It
will be forced magnetically into corotation with the neutron star, and part of
it will be ejected centrifugally, across the speed-of-light-cylinder (SLC)
distance, at transrelativistic speeds, corresponding to Lorentz
factors\ $\gamma$\ of a few. Note that pulsars are thought to boost their
electrons to Lorentz factors between $10^{3}$ and $10^{8}$ near the SLC, which
would amount to comparable or larger kinetic energies than those of the ions 
expelled by a GRBer.
Note also that a particle escaping (radially) from us with a Lorentz
factor\ $\gamma$\ of $5$ would be observed with a redshift of \ $z$
$\lesssim9$, according to equ.(1).

I therefore interpret the GRBs as the events when massive clumps
($\grtsim 10^{15}$g)\ from an inner accretion disk fall onto the surface of a
nearby neutron star, vaguely reminiscent of the accretion of comet
Shoemaker-Levy 9 by Jupiter in May 1994. More precisely, the short GRBs -- of
duration $\lesssim$ 2s -- are interpreted as the accretion events of one
single chunk. Their fading and softening early X-ray afterglows have shown
damped oscillations at the neutron star's spin period, between $5$s and $12$s,
for $\lesssim10^{2}$s, caused by periodic occultations of the impacted
neutron-star hemisphere, as familiar from the giant outbursts of the SGRs
(Gehrels et al 2006). Their lightcurve has the shape of a FRED (= fast rise,
exponential decay), whereby the spectrum softens during the decay, to be
understood as cooling. All the long bursts are superpositions of successive
short bursts (FREDs), with washed-out oscillations, and higher integrated
luminosities, cf. (Piran, 2004; Hjorth et al, 2006). Clearly, the long bursts
eject more matter than the short bursts do, so that SN-like lightcurves are
restricted to them. More massive chunks make wider FREDs than less massive ones.

Can this interpretation explain all the riddles (i) through (iv) listed at the
beginning of the last section? The almost isotropy of arrivals (i) has already
been explained as a (partial) compensation of column number densities by mild
beaming (in the disk planes) in (Kundt \& Chang 1993). Missing so far was a
good explanation for the large observed redshifts in the afterglows. They can
be understood by consultation of Figs.1b,2: When the impacted surface of the
neutron star flares, a GRB is emitted, viz. a wide-angle flash of hard
photons. Just ($\grtsim\Omega^{-1}$ $\approx$) one second later,
centrifugally ejected hot baryonic matter follows the photonic flash, as a
baryonic flash. Both flashes escape radially at high speeds, at \{ = ,
$\lesssim$\ \}the speed of light, impact on the circumstellar medium (CSM),
and cause it to radiate. A distant observer sees at first blue-shifted
radiation, from the near hemisphere of the CSM which is successively impacted
by the two expanding shells, for seconds after onsets, and subsequently from
fading, red-shifted radiation from the distant hemisphere, for hours and
months to come. Further light reaches the distant observer from CSM located
transverse to the line-of-sight, which dilutes the radiation from the front
and back side of the outgoing flashes. Still, both redshifted absorption and
redshifted emission (with the same z)\ are expected in general from the
baryonic shell when it is crossed by the photonic burst's
stimulated\ emission, and when it interacts directly with the CSM, with a
Lorentz factor $\gamma$ between 1 and 5. Which answers riddle (ii).

Riddle (iii) can be answered in a straight-forward manner: A significant
fraction of the two ejected shells' energies, photonic and baryonic, on
collision with their CSM, is expected to be radiated by the burst's CSM,
causing a strong light echo from the region which is impacted in $\grtsim$ a
light-crossing time, of radius $\lesssim$\ one light day during the first day
after the burst, and correspondingly for other times. This additional flaring,
or reflection nebula, can be mistaken for a distant host galaxy, with the same
redshift as the afterglow. In addition, occasional chance projections onto
luminous background nebulae cannot easily be ruled out, when only one or two
spectral lines are available for the identification.

And as concerns a similarity to a SN, riddle (iv), all we require is a dense
enough shell of ejecta from an energetic (long-lasting) burst of comparatively
small speed, $z$ $\lesssim$ $0.2$, corresponding to $\beta\lesssim$ $0.17$, in
which resonance scattering can store line photons for sufficiently long times
to cause exponentially declining lightcurves (instead of power laws), (Kundt
2008b). We deal with a phenomenon not too different from that of a SN explosion.

Remains a discussion of the twenty problems (for the cosmological
interpretation) listed at the end of the past section. Seventeen of them more
or less invite a local-Galactic (re-) interpretation, with frequent
repetitions (from additional impacts), and additional excitations (of the
dying pulsar). The problems are: (j) a resolved X-ray afterglow, (jj)\ pre-
and post-cursors to the bursts, (jjj) afterglows with long plateaus and abrupt
declines, (jv) superluminal expansions (for the assumed excessive distances),
(v) $z$-independence of the involved energetics, (vj) only mild distortions
(ionizations) of their CSM, (vjj) recorded flares from the impacted (patchy)
CSM, (vjjj) a non-cosmological energy dependence on $z$, (jx) spectral
excursions to $\lesssim10$ TeV, (x) occasional long durations, (xj) no
long-distance travel signatures, (xjj) no orphan afterglows (because of no
beaming), (xjjj) no detectable gravitational waves (because of $10^{16}$times
lower energetics), (xjv) no missing signals from the (expected) dead-pulsar
population, (xv) atypical hosts, (xvj) 2-temperature optical afterglows, and
(xvjj) a relation to the UHE CR generators, whose distances cannot be
cosmological, (because of the GZK\ cutoff).

The local-Galactic re-interpretation should likewise explain the remaining
three problems, viz.: (xvjjj) the cosmological-mimicking intervening damped
Lyman $\alpha$ (DLAs) and (xjx) metal-absorber systems, Mg II and C IV, apart
from distinct deviations of their distributions (Vreeswijk et al, 2004;
Sudilovsky et al , 2007): we may have to learn that the magnetars (or
whatever) can have unusually high (absorbing)\ column densities in their CSM.
And (xx) the famous `thin-shell' distribution of the GRBs by BATSE, inferred
from the log$N$-vs-log$S$ diagram \ -- \ and thickened a bit by the 
Pioneer-Venus-Orbiter data \ --
\ poses problems to both interpretations. In the local-Galactic
interpretation, it asks for a certain fine-tuning between the inhomogeneous
distribution of the Galaxy's nearby throttled pulsars and their somewhat
anisotropic emissions, which has been shown not to be prohibitive, though,
both analytically and numerically, by Kundt \& Chang (1993). When fundamental
physics clashes with circumstantial or statistical `evidence', my confidence
is in the former.

\section{Conclusions}

Mainstream interpretations of GRBs struggle with intensity factors of
$10^{16}$ when compared with the one by Galactic neutron stars, because of a
distance ratio of $10$Gpc/$10^{2}$pc = $10^{8}$; which shrinks (only) to
$10^{10}$ when a (large!) beaming factor of $10^{3}$ is assumed, based on
several poorly understood mechanisms. The problem disappears when it is
realized that redshifts need not mean distances. Note that the problem exists
for all the emissions: not only for the prompt, hard emissions, but likewise
for the (often similarly hard) afterglow emissions, from radio all the way up
through (early!) optical to X-ray brightnesses, and even occasionally to TeV
energies (with dominating power!).

\begin{acknowledgements}

I wish to thank G\"{u}nter Lay and Ole Marggraf for repeated help with the
electronic data handling, both in the preparation of the power-point
presentation, and in its present documentation, Gernot Thuma and Hans 
Baumann for the MS, and Ingo Thies for the submission to astro-ph.

\end{acknowledgements}

\bibliographystyle{aa}

\end{document}